\documentclass[showpacs,preprintnumbers,
superscriptaddress,amsmath]{revtex4}

\usepackage{epsfig}

\begin{document}
\preprint{\vtop{\hbox{RU06-1-B}\hbox{MCTP-02-06}
\vskip24pt}}

\title{Absence of the London limit for the first-order phase
transition to a color superconductor}

\author{Jorge L.\ Noronha}
\email{j.noronha@figss.uni-frankfurt.de} \affiliation{Frankfurt
Institute for Advanced Studies, J.W.\ Goethe--Universit\"at, D-60438
Frankfurt am Main, Germany}
\author{Hai-cang Ren}
\email{ren@summit.rockefeller.edu} \affiliation{Physics Department,
The Rockefeller University, 1230 York Avenue, New York, New York
10021-6399, USA} \affiliation{Institute of Particle Physics, Central
China Normal University, Wuhan, 430079, China}
\author{Ioannis Giannakis}
\email{giannak@summit.rockefeller.edu} \affiliation{Physics
Department, The Rockefeller University, 1230 York Avenue, New York,
New York 10021-6399, USA}
\author{Defu Hou}
\email{hdf@iopp.ccnu.edu.cn} \affiliation{Institute of Particle
Physics, Central China Normal University, Wuhan, 430079, China}
\author{Dirk H.\ Rischke}
\email{drischke@th.physik.uni-frankfurt.de} \affiliation{Frankfurt
Institute for Advanced Studies, J.W.\ Goethe--Universit\"at, D-60438
Frankfurt am Main, Germany} \affiliation{Institut
f\"ur Theoretische Physik, J.W. Goethe--Universit\"at, D-60438
Frankfurt am Main, Germany}

\begin{abstract}

We study the effects of gauge-field
fluctuations on the free energy of a homogeneous color
superconductor in the color-flavor-locked (CFL) phase.
Gluonic fluctuations induce a strong 
first-order phase transition, in contrast to electronic 
superconductors where this transition is weakly first order.
The critical temperature for this transition
is larger than the one corresponding to the diquark pairing
instability. The physical reason is that the gluonic Meissner masses 
suppress long-wavelength fluctuations as compared to the normal
conducting phase where gluons are massless, which stabilizes the
superconducting phase.
In weak coupling, we analytically compute the 
temperatures associated with the limits of metastability of the
normal and superconducting phases, as well as 
the latent heat associated with the first-order
phase transition. We then extrapolate our results to intermediate densities
and numerically evaluate the temperature of the fluctuation-induced
first-order phase transition, as well as the discontinuity of the diquark
condensate at the critical point.  We find that
the London limit of magnetic interactions is absent in color superconductivity.

\end{abstract}

 \maketitle

\section {Introduction}

It was suggested a long time ago that quark matter might exist
within the central regions of superdense stars \cite{itoh}. Since
Quantum Chromodynamics (QCD) is an asymptotically free theory
\cite{asymptfree}, it was argued that the extremely compressed
matter found in neutron stars consists of quarks rather than of
hadrons and that realistic calculations in the framework of
QCD become possible \cite{collinsperry}. At high baryon
densities and sufficiently low temperatures, however, a phase
transition between normal and color-superconducting quark matter is
expected \cite{love, reviews, dirkreview, recentreviews}. Therefore,
color superconductivity (CSC) may be relevant to explain several
important aspects of the highly compressed matter present in compact
stars, e.g., the cooling rates \cite{astrostuff}, and the rotational
properties of stars \cite{madsen}. Nevertheless, trustworthy
perturbative calculations can only be performed for ultra-high
chemical potentials, $\mu \gg \Lambda_{\rm QCD}$. The weak-coupling
expansion of the temperature for the diquark pairing instability
reads \cite{son, gapTc}
\begin{equation}
\ln\frac{T_c}{\mu}=-\frac{3\pi^2}{\sqrt{2}g}
+\ln\frac{2048\sqrt{2}\pi^3}{9\sqrt{3}g^5}
+\gamma-\frac{\pi^2+4}{8}+O(g)\;,
\label{pairing}
\end{equation}
where $g$ is QCD running coupling constant at the chemical potential
$\mu$, and $\gamma$ is the Euler-Mascheroni constant.
The energy gap at $T=0$ in the CFL phase is
\cite{dirkreview}
\begin{equation}
\Delta(0)=2^{-\frac{1}{3}}\pi e^{-\gamma}\,T_c\;.
\label{gap}
\end{equation}
It follows from the
Ginzburg-Landau (GL) theory of CSC in weak coupling that the phase
transition is of second order at $T_c$ and the GL
parameter is \cite{gia}
\begin{equation}
\kappa=\sqrt{\frac{72\pi^3}{7\zeta(3)\alpha_s}}\,\frac{T_c}{\mu}\;,
\label{glp}
\end{equation}
with $\alpha_s=g^2/(4\pi)$. When $g\rightarrow 0$ then $\kappa
\rightarrow 0$ and, therefore, the CFL color superconductor is of
extreme type I.

It is well known in the context of electronic superconductors that
gauge-field fluctuations change the second-order phase transition
into a first-order transition \cite{ma}. However, the strength of this
first-order transition is sensitive to the relationship among the
three length scales that are involved: the coherence length near the
transition, $\xi$, the magnetic penetration depth near the
transition, $\lambda$, and the coherence length at $T=0$,
$\xi_0\sim\frac{1}{T_c}$. A superconductor with $\lambda \gg \xi_0$
is said to be in the London limit. In this case, the coupling between the
gauge field and the order parameter is approximately local. The
opposite case, $\lambda \ll \xi_0$, corresponds to the Pippard limit
and the coupling becomes highly nonlocal \cite{lif}. For a type-I
electronic superconductor the Pippard limit is always realized at
$T=0$. As the temperature is raised towards the transition to
normal quark matter, the penetration depth increases and so does
the ratio $\lambda/\xi_0$. A crossover from the Pippard limit to the
London limit would be expected if the transition is of second order.
How does the first-order phase transition induced by gauge-field
fluctuations change this scenario? Are both limits still realized?
In the case of known electronic superconductors of type I, the
first-order phase transition is sufficiently weak to warrant a
crossover between Pippard and London limits,
which has been indeed observed experimentally for strong
type-I materials like aluminum \cite{bonalde}. However, the
situation is completely different for a color superconductor. It was
shown in a previous work \cite{dirkfluct} that $\lambda\ll\xi_0$ is
maintained at the phase transition for asymptotically high baryon
density. As will be shown below, this feature remains valid when the
results of Ref.\ \cite{dirkfluct} are extrapolated to moderately
high densities.

The current work, which is an 
extension and continuation of the previous project
\cite{dirkfluct}, is organized as follows. In the next section, we
shall review the generalized GL free energy derived
previously. The relevant thermodynamic quantities of the
first-order color-superconducting transition will be calculated in
weak coupling in Sec.\ \ref{III} and the extrapolation of the results to
moderate coupling will be presented in Sec.\ \ref{IV}. Concluding
remarks will then be given in Sec.\ \ref{V}. Moreover, technical details
on the derivation of the generalized GL free energy, which were
skipped in Ref.\ \cite{dirkfluct}, will be sketched in the appendix.
In contrast to Ref.\ \cite{dirkfluct},
the zero-temperature coherence length will be
defined as $\xi_0=1/(2\pi T_c)$. Our units are
$\hbar=c=k_B=1$ and 4-vectors are denoted by capital letters,
$K \equiv K^\mu = (\omega, \vec{k})$. In our formulas, Tr 
indicates the summation over all indices
including momentum, $\vec{k}$, and energy, $\omega$,
while tr denotes the summation over
all indices except momentum and energy.

\section {The generalized Ginzburg-Landau free energy} \label{II}

\begin{figure}[t]
\includegraphics[width=14cm]{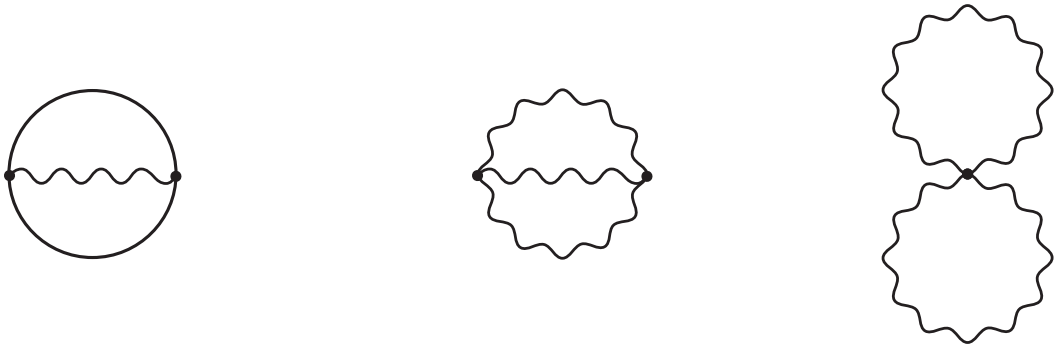}
\caption{The two-loop approximation to $\Gamma_2$. Straight lines
denote quark propagators, wavy lines denote gluon propagators.}
\label{gamma2}
\end{figure}

The CJT effective potential \cite{dirkreview} reads
\begin{equation} \label{CJTpot}
\Gamma[\bar{\cal D},\bar{\cal S}] = 
\frac{T}{2 \, \Omega} \left\{ {\rm Tr} \ln \bar{\cal D}^{-1}
+ {\rm Tr} (D^{-1} \bar{\cal D} -1) - 
{\rm Tr} \ln \bar{\cal S}^{-1}
- {\rm Tr} (S^{-1} \bar{\cal S} -1)   -
 2 \, \Gamma_2[\bar{\cal D},\bar{\cal S}]\right\}
\;,
\end{equation}
where $\Omega$ denotes the 3-volume of the system, 
$\bar{\cal D}$ and $\bar{\cal S}$ are the full gluon and quark
propagators, $D^{-1}$ and $S^{-1}$ are the corresponding inverse tree-level
propagators, and $\Gamma_2$ is the sum of all two-particle irreducible vacuum diagrams.
We work in the two-loop approximation, i.e.,
$\Gamma_2$ contains only the diagrams shown in Fig.\ \ref{gamma2}.
The first diagram, containing quark propagators, leads to
a term of order $g^2 \mu^4 $ in $\Gamma$, while the other two
diagrams, containing only gluon propagators, lead to terms
proportional to powers of $T$.
Therefore, at small temperatures $T \sim T_c \sim \mu \exp(-1/g)$,
one can drop the last two
diagrams and restrict the consideration to the first.
In explicit form,
\begin{equation} \label{Gamma2}
\Gamma_2 [\bar{\cal D},\bar{\cal S}] =
- \frac{1}{2}\,{\rm Tr} \{ \bar{\cal D}\, \Pi[\bar{\cal S}] \}\;,
\end{equation}
where 
\begin{equation} \label{Pi}
\Pi[\bar{\cal S}] \equiv 
\frac{1}{2} \,{\rm Tr} ( \hat{\Gamma}\, \bar{\cal S}\, \hat{\Gamma}\,
\bar{\cal S})
\end{equation}
is a functional of the full quark
propagator $\bar{\cal S}$ and the bare quark-gluon vertex $\hat{\Gamma}$.
Note that the trace in Eq.\ (\ref{Gamma2}) is over momenta, as well as Lorentz and
adjoint color indices, while in Eq.\ (\ref{Pi}) it is over momenta, as well
Dirac, flavor, and fundamental color indices. The minus sign in
Eq.\ (\ref{Gamma2}) takes account of the fermion loop and the factor
1/2 is due to the fact that this is a second-order correction
to the CJT effective potential.
The factor 1/2 in Eq.\ (\ref{Pi}) accounts for the doubling of the
fermionic degrees of freedom in Nambu-Gor'kov space.

The free energy is given by the CJT effective potential at
its stationary points, determined by
\begin{equation}
\left. \frac{\delta \Gamma}{\delta \bar{\cal D}}\right|_{\bar{\cal D} = {\cal
D}, \bar{\cal S} = {\cal S}}= 0 \;, \;\;\;\;
\left. \frac{\delta \Gamma}{\delta \bar{\cal S}}\right|_{\bar{\cal D} = {\cal
D}, \bar{\cal S} = {\cal S}} = 0\;.
\end{equation}
The first condition gives a Dyson-Schwinger equation for the
gluon propagator,
\begin{equation}
{\cal D}^{-1} = D^{-1}+ \Pi[{\cal S}]\; .
\end{equation}
One now sees that, at the stationary point, the second term
in Eq.\ (\ref{CJTpot}) cancels the last term, 
\begin{equation} \label{CJTpot2}
\Gamma[{\cal D},{\cal S}]
= \frac{T}{2 \Omega} \left[ {\rm Tr} \ln {\cal D}^{-1}
 -  {\rm Tr} \ln {\cal S}^{-1}
- {\rm Tr} (S^{-1} {\cal S} -1) \right]\;.
\end{equation}
In terms of the gluon and quark propagators in the normal phase, ${\cal
D}_n(K)$ and ${\cal S}_n(K)$, the propagators in the superconducting
phase are written as
\begin{subequations} \label{eqnick}
\begin{eqnarray}
{\cal S}(K)& =& {\cal S}_n(K)+\delta{\cal S}(K,\Delta) \; , \\
{\cal D}^{-1}(K)& =& {\cal D}_n^{-1}(K)+\delta\Pi(K,\Delta)\;,
\end{eqnarray}
\end{subequations}
where $\delta \Pi \equiv \Pi - \Pi_n$, i.e.,
${\cal D}_n^{-1}$ {\em already\/} contains the 
hard-dense-loop (HDL) resummed gluon self-energy $\Pi_n$.
The gluon self-energy in the superconducting phase, $\Pi$,
depends on the superconducting gap parameter, $\Delta$, and so
$\delta \Pi$ also depends on $\Delta$.
Similarly, the quark propagator in the normal phase
${\cal S}_n$ contains quark self-energy corrections,
and $\delta {\cal S}$ depends on $\Delta$.

Inserting Eqs.\ (\ref{eqnick}) into Eq.\ (\ref{CJTpot2}),
we obtain
\begin{equation}
\Gamma=\Gamma_n+{\Gamma}_{\rm cond}+{\Gamma}_{\rm fluc}+
\Gamma_{\rm fluc}^\prime\;,
\label{eqerie}
\end{equation}
where
\begin{subequations}
\begin{eqnarray} \label{gamman}
\Gamma_n & = & \frac{T}{2\, \Omega} \left[ {\rm Tr} \ln {\cal D}_n^{-1}
 -  {\rm Tr} \ln {\cal S}_n^{-1}
- {\rm Tr} (S^{-1} {\cal S}_n -1) \right]\;, \\
{\Gamma_{\rm cond}}& =& \frac{T}{2\, \Omega}
\left[ {\rm Tr} ({\cal D}_n{\delta}{\Pi})
-{\rm Tr}( S^{-1}\delta{\cal S}) +
{\rm Tr}\ln(1+{\cal S}_n^{-1}\delta{\cal S}) \right]\;,
\label{eqbp}\\
{\Gamma}_{\rm fluc}& =& \frac{T}{2\,\Omega}
{\sum_{\vec k, \omega=0}} {\rm tr} \left\{
\ln \left[ 1+{\cal D}_n(K){\delta}{\Pi}(K,\Delta)\right]
 -  {\cal D}_n(K){\delta}{\Pi}(K,\Delta) \right\}\; ,
\label{eqmunif} \\
{\Gamma}_{\rm fluc}^\prime & =& \frac{T}{2\,\Omega}
{\sum_{\vec k, \omega\neq 0}} {\rm tr} \left\{
\ln\left[ 1+{\cal D}_n(K){\delta}{\Pi}(K,\Delta) \right]
 - {\cal D}_n(K){\delta}{\Pi}(K,\Delta) \right\}\;.
\label{eqmunifprime}
\end{eqnarray}
\end{subequations}
The generalized GL free energy is the {\rm difference\/} in the CJT
effective potential between the superconducting phase and
the normal phase, $\Gamma - \Gamma_n$. It includes both
the ordinary GL terms and the fluctuation terms. 
Note that we have added a term ${\rm Tr} ({\cal D}_n \delta \Pi)$ in
$\Gamma_{\rm cond}$ and simultaneously subtracted it in
$\Gamma_{\rm fluc}, \Gamma_{\rm fluc}'$. This term corresponds
to the so-called exchange (free) energy \cite{kapusta} and {\em
must\/} be present in order to obtain the correct
expression for $\Gamma_{\rm cond}$ \cite{Iida,ioannis}.
Therefore, we {\em have\/} to subtract it in the fluctuation part of the
free energy. Only with this subtraction, 
$\Gamma_{\rm fluc} + \Gamma_{\rm fluc}'$ represents the well-known plasmon ring
resummation \cite{kapusta}. We note in passing that it is quite
gratifying to see that the CJT formalism naturally contains
all these different many-body contributions to the free energy.

In Ref.\ \cite{baym}, the exchange energy was not
subtracted from the plasmon ring contribution.
This leads to an overall change of sign of the fluctuation energy.
As shown below [see Eq.\ (\ref{eqrigas})], the contribution
of the fluctuation energy is 
$\sim \ln (1+u) -u$, which is always negative, while 
in Ref.\ \cite{baym} it is $\sim \ln (1+u)$ which is positive (for
$u>0$). Therefore, the authors of Ref.\ \cite{baym} concluded that
gauge-field fluctuations raise the free energy of the
color-superconducting phase, and thus decrease the transition
temperature to the normal phase. In our case, however, 
the gauge-field fluctuations decrease the free energy, i.e., 
stabilize the color-superconducting phase and therefore lead
to a larger transition temperature. 

This is physically plausible if one remembers that
gauge-field fluctuations are also present in the normal phase,
namely in the first term in Eq.\ (\ref{gamman}).
Since transverse gluons are massless in the normal phase, $\Pi_n(0) = 0$,
long-wavelength fluctuations are enhanced over those
in the color-superconducting phase where gluons are massive, $\delta
\Pi \neq 0$.
Thus, the fluctuation energy in the normal phase is larger than
in the superconducting phase.

As shown in the appendix, the weak-coupling approximation gives
rise to \cite{Iida,ioannis}
\begin{equation}
\Gamma_{\rm cond} =  \frac{6\mu^2}{\pi^2}\,t \, \Delta^2(T)
+\frac{21\zeta(3)}{4\pi^4}\left(\frac{\mu}{T_c}\right)^2\Delta^4(T),
\label{free}
\end{equation}
where $T_c$ is determined up to the accuracy of Eq.\ (\ref{pairing})
and $t\equiv (T-T_c)/T_c$ is the reduced temperature. 
The gap parameter of the fermionic
quasiparticle excitations is $\Delta$ (8-fold) and $2\Delta$
(1-fold) \cite{rischke}. One can check that the quadratic and
quartic coefficients of $\Gamma_{\rm cond}$ for CSC are,
respectively, 12 ($=8\times 1^2+2^2$) and 24 ($=8\times 1^4+2^4$)
times larger than those for an electronic superconductor. The
relevant fluctuation term is
\begin{equation}
\label{fluct} \Gamma_{\rm fluc}=8\,T\int\frac{d^3\vec
k}{(2\pi)^3}\left\{ \ln\left[1+
\frac{m^2(T,k)}{k^2}\right]-\frac{m^2(T,k)}{k^2}\right\},
\end{equation}
while $\Gamma_{\rm fluc}^\prime$ is of higher order. The
momentum-dependent Meissner mass reads
\begin{equation}
m^2(T,k)=\frac{1}{\lambda^2}f(k\xi_0),
\end{equation}
with the chromomagnetic penetration depth given by
\begin{equation}
\frac{1}{\lambda^2}
=\frac{7\zeta(3)}{24\pi^4}\frac{g^2\mu^2\Delta^2}{T_c^2},
\end{equation}
and
\begin{equation}
f(y)=\frac{6}{7{\zeta}(3)}{\sum_{s=0}^{\infty}}{\int^{1}_0}dx
\frac{1-x^2}{(s+{\frac{1}{2}})[4(s+{\frac{1}{2}})^2+y^2x^2]}\;.
\label{eqkirkos}
\end{equation}
Carrying out the integration in Eq.\ (\ref{fluct}) and combining the
result with Eq.\ (\ref{free}), we find
\begin{equation}
\Gamma =  \frac{6\mu^2}{\pi^2}\,t \, \Delta^2(T)
+\frac{21\zeta(3)}{4\pi^4}\left(\frac{\mu}
{T_c}\right)^2\Delta^4(T)
+ 32\pi(T_c)^4\, F\left(\frac{\xi_0^2}{\lambda^2(T)}\right),
\label{ggl}
\end{equation}
where the function $F$ is defined by
\begin{equation}
F(z)={\int^{\infty}_0}dx\, x^2 \left\{ \ln \left[1+{\frac{z}{x^2}}f(x)\right]
-\frac{z}{x^2}f(x) \right\}\;.
\label{eqrigas}
\end{equation}
The London limit corresponds to small arguments in Eqs.\
(\ref{eqkirkos}) and (\ref{eqrigas}). We have
\begin{equation}
f(y) = 1 - \frac{31}{140}\frac{\zeta(5)}{\zeta(3)}y^2 + O(y^4),
\label{london1}
\end{equation}
and
\begin{equation}
F(z) \simeq -\frac{\pi}{3}z^{\frac{3}{2}}. \label{london2}
\end{equation}
In the Pippard limit the arguments of $f(y)$ and $F(z)$ become large
and we end up with
\begin{equation}
f(y)= \frac{3\pi^3}{28\zeta(3)y}\left[1-12\, \frac{4\ln y+4\ln
2+\gamma}{\pi^3 y} +O(y^{-2})\right] \label{pippard1}
\end{equation}
and
\begin{equation}
F(z) \simeq  -\frac{3\pi^3}{28\zeta(3)}z
\Big[\ln\Big(\frac{3\pi^3}{28\zeta(3)}z\Big)+{\rm const}\Big].
\label{pippard2}
\end{equation}
Here we have retained the first corrections for both limits of the
function $f(y)$ in order to assess the deviation from each limit at
the CSC phase transition.

Before concluding this section, let us clarify once more several
differences between our formulation and that of Ref.\ \cite{baym}.
First, since in their treatment the term $-m^2(T,k)/k^2$ 
in Eq.\ (\ref{fluct}), which arises from the subtraction of the 
exchange energy, is missing, their formal power-series
expansion for the fluctuation energy in terms of $\Delta$ starts already
at quadratic order. This then gives rise to a renormalized 
critical temperature $T_c'$.
When including the term in question, there is no
such renormalization of $T_c$. 
Nevertheless, the difference between $T_c$ and $T_c'$ is
of order $\mathcal{O}(g^2)$, and the two-loop approximation 
employed in the derivation of Eq.\ (\ref{ggl}) is not sufficiently accurate
to provide all corrections of this order. 
Furthermore, the authors of Ref.\ \cite{baym} approximated the
momentum-dependent Meissner mass by a constant and simply cut off the
momentum integration in Eq.\ (\ref{fluct}).
In that case, one is effectively in the London limit
of Eqs.\ (\ref{eqkirkos}) and (\ref{eqrigas}), where the
fluctuation energy is of the form of Eq.\ (\ref{london2}). 
The shift in the transition temperature compared
to that of Eq.\ (\ref{pairing}) is then only of order
$\mathcal{O}(g^2)$ \cite{baym}, and not of order
$\mathcal{O}(g)$, as found here and in Ref.\ \cite{dirkfluct}.

\section {The first-order CSC transition in weak coupling} \label{III}

A generic first-order phase transition can be described by three
characteristic temperatures: the transition temperature $T_c^*$, 
the maximum temperature of the (metastable) superheated superphase 
$T_{\rm sh}$, and the minimum temperature of the (metastable)
supercooled normal phase $T_{\rm sc}$, respectively.
These temperatures are related in the following way,
\begin{equation}
T_{\rm sc} < T_c^* < T_{\rm sh}\;,
\end{equation}
and they can be obtained from the generalized GL free
energy (\ref{ggl}). The lower margin of a supercooled normal phase
corresponds to
\begin{equation}
\left. \frac{\partial^2\Gamma}{\partial\Delta^2}\right|_{\Delta=0}=0,
\end{equation}
and, using Eq.\ (\ref{ggl}), we have
\begin{equation}
T_{\rm sc}=T_c\; ,
\end{equation}
which relates $T_{\rm sc}$ with the onset temperature for diquark
pairing. On the other hand, the transition occurs at 
\begin{equation}
\frac{\partial\Gamma}{\partial\Delta}=0\;, 
\qquad \Gamma = 0\;,
\end{equation}
for a value of $\Delta \equiv \Delta_{c^*} \neq 0$.
This implies that
\begin{eqnarray}
t_c^* & + &
\frac{7\zeta(3)}{4\pi^2}\frac{\Delta^{2}_{c^*}}{T^{2}_c}+
\frac{7\zeta(3)}{18\pi^3}g^2F^{\prime}\left(\frac{\xi_0^2}{\lambda^2_{c^*}}
\right)=0\;, \nonumber \\
t_c^* & + &
\frac{7\zeta(3)}{8\pi^2}\frac{\Delta^{2}_{c^*}}{T^{2}_c}+
\frac{7\zeta(3)}{18\pi^3}g^2\frac{\lambda^2_{c^*}}{\xi_0^2}
F\left(\frac{\xi_0^2}{\lambda^2_{c^*}}\right)=0\;.
\label{transition}
\end{eqnarray}
Eliminating $t_c^*$ in the equations above we have
\begin{equation}
{\cal F}\left(\frac{\xi_0^2}{\lambda^2_{c^*}}\right)=
\frac{216{\pi}^7}{7{\zeta}(3)g^4}\left(\frac{T_c}{\mu}\right)^2\;,
\label{eqvrema}
\end{equation}
where ${\cal F}(z)=-F^{\prime}(z)/z+F(z)/z^2$. Solving Eq.\
(\ref{eqvrema}) for $\Delta^{2}_{c^*}$, with the aid of Eq.\
(\ref{pippard2}), we obtain
\begin{equation}
\Delta^{2}_{c^*}=\frac{\pi^2}{63{\zeta}(3)}\,g^2T^{2}_c\;.
\label{deltac}
\end{equation}
The transition temperature is obtained substituting Eq.\
(\ref{deltac}) into either one of Eqs.\ (\ref{transition}), which
produces
\begin{equation}
T_c^*=\Big(1+\frac{\pi^2}{12\sqrt{2}}g\Big)\,T_c\;.
\label{weak}
\end{equation}
These were the results reported in Ref.\ \cite{dirkfluct}. The
penetration depth at the transition is
\begin{equation}
\frac{1}{\lambda^2_{c^*}}=\frac{g^4}{216\pi^2}\mu^2\;,
\end{equation}
which yields the ratio
\begin{equation}
\frac{\xi_0^2}{\lambda^2_{c^*}}=\frac{g^4\mu^2}{864\pi^4T_c^2}\gg1\;.
\end{equation}
Thus, the Pippard limit is valid for the entire CSC phase at
sufficiently large chemical potentials.

We shall proceed to determine $T_{\rm sh}$. The free energy $\Gamma$
as a function of $\Delta$ has a local maximum between $\Delta=0$
and the minimum $\Delta_{c^*}$ at $T=T_c^*$ in the superconducting
phase. As $T$
increases, the local minimum remains unchanged until it coalesces
with the local maximum, where
\begin{equation}
\frac{\partial\Gamma}{\partial\Delta}=0\;, \qquad
\frac{\partial^2\Gamma}{\partial\Delta^2}=0\;,
\end{equation}
for a value of $\Delta\equiv \Delta_{\rm sh} \neq 0$.
It then follows that
\begin{eqnarray}
t_{\rm sh} + \frac{7{\zeta}(3)}{4\pi^2}\frac{{\Delta^{2}_{\rm sh}}}
{T^{2}_c} &+&
\frac{7{\zeta}(3)}{18\pi^3}g^2F^{\prime}\left(\frac{\xi_{0}^2}
{\lambda_{\rm sh}^2}\right)=0\;, \nonumber \\
F^{\prime\prime}\left(\frac{\xi_{0}^2}{\lambda_{\rm sh}^2}\right)
&=& -\frac{432\pi^7}{7\zeta(3)g^4}\Big(\frac{T_c}{\mu}\Big)^2\;.
\label{overheat}
\end{eqnarray}
Moreover, Eq.\ (\ref{overheat}) together with Eq.\ (\ref{pippard2})
yields
\begin{equation}
\Delta^2_{\rm sh}=\frac{\pi^2}{126\zeta(3)}g^2T^{2}_{c^*}
=\frac{1}{2}\Delta^{2}_{c^*}\;.
\end{equation}
Subtracting the first equation in Eq.\ (\ref{transition}) from Eq.\
(\ref{overheat}) and using (\ref{pippard2}), we find that
\begin{equation}
t_{\rm sh}-t_c^*=\frac{g^2}{72}(1-\ln{2})\;,
\end{equation}
and as a result
\begin{equation}
T_{\rm sh}=\left[1+\frac{g^2}{72}(1-\ln{2})\right]T_c^*\;.
\end{equation}
Note that $T_c^*$ is one order of $g$ closer to $T_{\rm sh}$ than
to $T_{\rm sc}$. The ratio
\begin{equation}
\frac{\xi_0^2}{\lambda_{\rm sh}^2}=\frac{g^4\mu^2}{1728\pi^4T_c^2}
\end{equation}
implies that even the metastable CSC state is in the Pippard limit
in weak coupling. Although the diagrammatics behind the generalized
GL free energy function (\ref{ggl}) determine $T_c$
only up to subleading order, the leading-order differences among
the three characteristic temperatures do not change if higher-order
corrections to $T_c$ are included.

Another observable associated with the first-order phase transition
is the latent heat $L=T_c{\Delta}S$, where $\Delta S$ is the change
in entropy density at the transition. We have
\begin{equation}
\Delta S=-\Big ( \frac{\partial\Gamma}{\partial T} \Big
)_{\Delta=\Delta_c^*} =\frac{2g^2}{21\zeta(3)}{\mu^2}T_c\;,
\end{equation}
and as a result
\begin{equation}
L=\frac{2g^2}{21\zeta(3)}{\mu^2}T^{2}_c
\equiv \frac{6 \mu^2}{\pi^2} \, \Delta_{c^*}^2\;.
\end{equation}
Now we calculate the strength of the first-order phase transition as
was defined in Ref.\ \cite{ma},
\begin{equation}
t_{\rm HLM}=\frac{L}{{\Delta}c_v}\;,
\end{equation}
where ${\Delta}c_v$ is the jump in specific heat at the second-order
phase transition, ignoring the fluctuations. If we ignore the third
term in Eq.\ (\ref{ggl}) we recover the ordinary GL theory from
which we find $\Delta c_v=24\mu^{2}T_c/[7\zeta (3)]$. Thus, we
have
\begin{equation}
\frac{t_{\rm HLM}}{T_c}=\frac{g^2}{36}\;. \label{strength}
\end{equation}
Note that Eq.\ (\ref{strength}) implies that the strength of the
first-order phase transition weakens (logarithmically) with increasing chemical
potential, which is in agreement with the fact that the
second-order phase transition is recovered at asymptotically large
densities. Note that for electronic superconductors,
$t_{\rm HLM}/T_c \sim 10^{-6}$ \cite{ma} which, for realistic
values of $g \sim 1$ is much smaller
than the right-hand side of Eq.\ (\ref{strength}).

\section {Numerical results}\label{IV}

Strictly speaking, the weak-coupling results in the previous section
are only valid at ultra-high baryon densities such that
$\mu\gg\Lambda_{\rm QCD}$. For quark matter that may exist inside a
compact star $\mu$ is expected to be slightly higher than
$\Lambda_{\rm QCD}$, and then the weak-coupling expansion becomes
problematic. Nevertheless, we shall assume that the generalized
GL free energy remains numerically reliable down to
realistic quark densities. Even though this is not the case, the
qualitative statement for the absence of the London limit in CSC may
still survive, due to the reason given at the end of this section.


\begin{figure}[t]
\includegraphics[width=11.0cm,angle=-90]{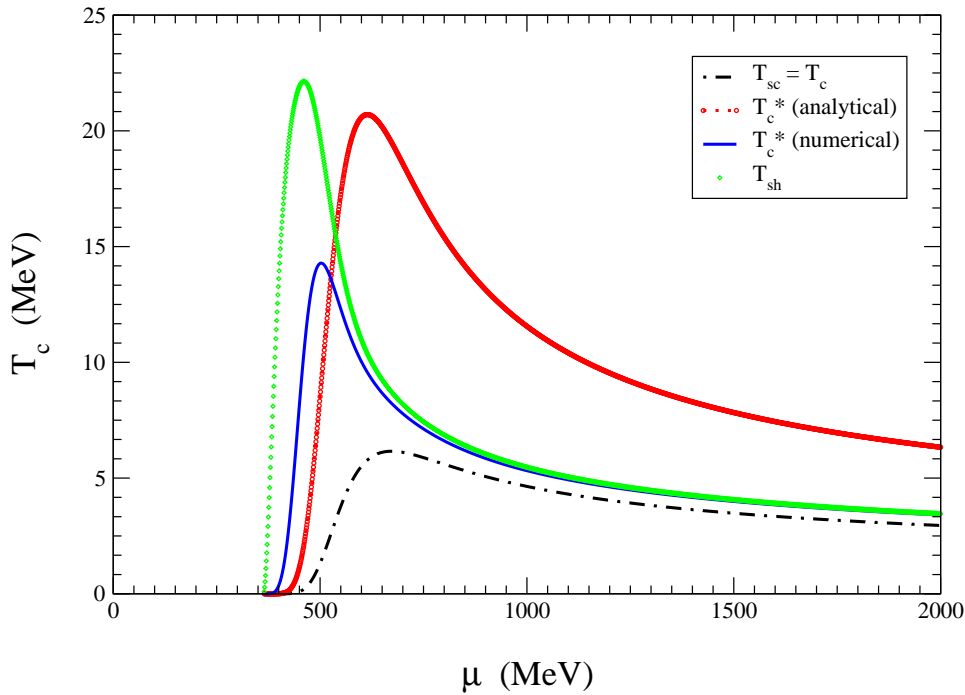}
\caption{Comparison between the different temperatures involved in
the discussion of the fluctuation-induced first-order phase
transition.} \label{fig1}
\end{figure}

We solved Eq.\ (\ref{eqvrema}) and the second equation in Eq.\
(\ref{overheat}) numerically in order to find $\Delta_{c^*}$ and
$\Delta_{\rm sh}$ as functions of the chemical potential. The
transition temperature $T_c^*$ is obtained using $\Delta_{c^*}$ in
either one of Eqs.\ (\ref{transition}) and the temperature $T_{\rm
sh}$ is obtained from the first equation in Eq.\ (\ref{overheat}).
We use the 3-loop formula for $\alpha_s=g^2/4\pi$ \cite{pdg},

\begin{eqnarray}
\alpha_s(\mu) &=& \frac{4\pi}{\beta_{0}\ln(\mu^2/\Lambda_{\rm
QCD}^2)}\left[1-\frac{2\beta_1}
{\beta_{0}^2}\frac{\ln[\ln(\mu^2/\Lambda_{\rm
QCD}^2)]}{\ln(\mu^2/\Lambda_{\rm QCD}^2)}\right.
\nonumber \\
&+& \left. \frac{4\beta_{1}^2}{\beta_{0}^4[\ln(\mu^2/\Lambda_{\rm
QCD}^2)]^2} \left( \left\{\ln\left[\ln\left(\frac{\mu^2}{
\Lambda_{\rm QCD}^2}\right)\right]-1/2\right\}^2
+\frac{\beta_{2}\beta_{0}}{8\beta_{1}^{2}} -\frac{5}{4}    \right)
\right],
\end{eqnarray}
where $\beta_0 = 9$, $\beta_1=51-19/(3N_f)=32$,
$\beta_2=2857-5033N_f/9+ 325N_{f}^{2}/27$, for three
colors and three flavors $N_f=3$. Moreover, we have taken
$\Lambda_{QCD}=364$ MeV in our calculations, in order to obtain
the correct value of $\alpha_s$ at the scale of the $Z$-boson mass.

Figure \ref{fig1} shows the three temperatures $T_{\rm sc}$, $T_c^*$ and
$T_{\rm sh}$ as functions of the chemical potential, along with the
weak-coupling formula (\ref{weak}) obtained in Ref.\
\cite{dirkfluct}. Note that $T_c^*$ is still closer to $T_{\rm sh}$
than to $T_{\rm sc}$ down to few hundreds of MeV.
\begin{figure}[t]
\includegraphics[width=11.0cm,angle=-90]{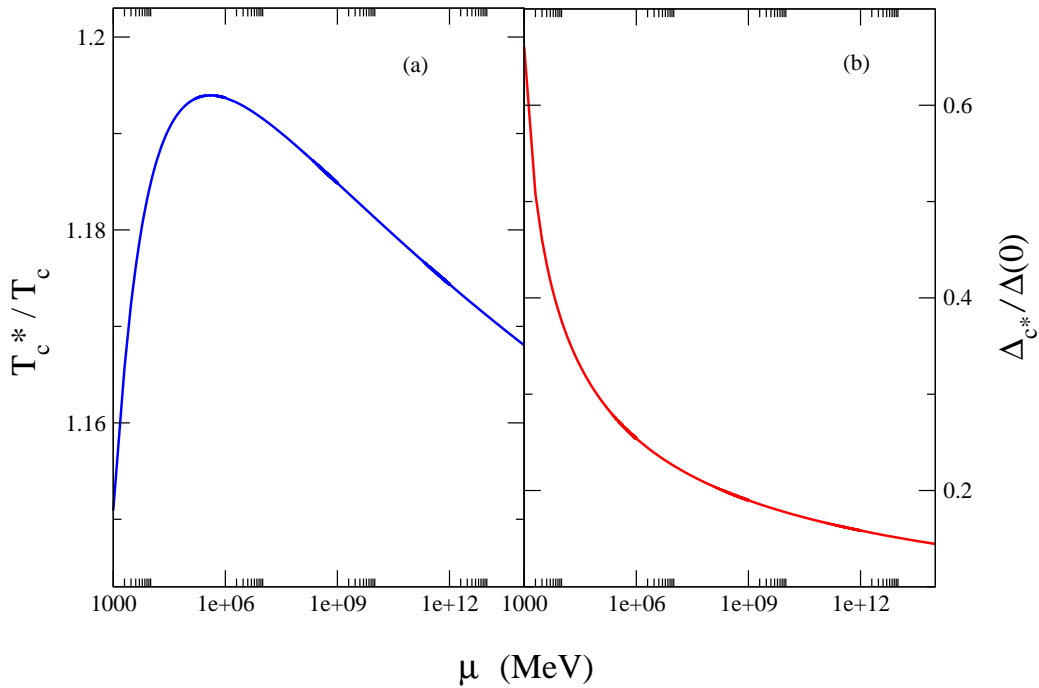}
\caption{(a) Comparison between the critical temperatures at high
densities. (b) Discontinuity of the gap at the transition.}
\label{fig2}
\end{figure}
A comparison between the numerically evaluated critical temperature
$T_c^*$ and $T_c$ is shown in Fig.\ \ref{fig2} (a) and the discontinuity of
the gap at $T_c^*$, relative to its value at $T=0$, (\ref{gap}), is
shown in Fig.\ \ref{fig2} (b) \cite{shortpaper}. Both plots indicate that
\begin{equation}
\lim_{\mu\to\infty}\frac{T_c^*}{T_c}=1\;, \qquad
\lim_{\mu\to\infty}\frac{\Delta_{c^*}}{\Delta(0)}=0\;,
\end{equation}
as expected, although the convergence is rather slow.

\begin{figure}[h]
\includegraphics[width=11.0cm,angle=-90]{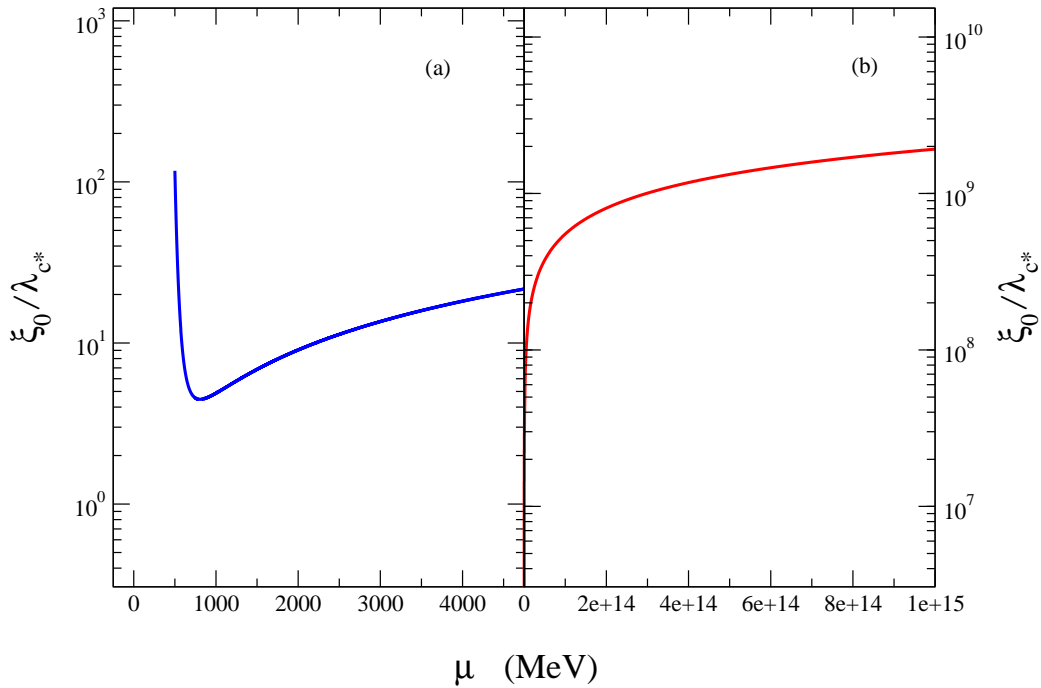}
\caption{(a) $\xi_0/\lambda_{c^*}$ at small chemical
potentials. (b) The same ratio at very large chemical potentials.}
\label{fig3}
\end{figure}

\begin{figure}[h]
\includegraphics[width=11.0cm,angle=-90]{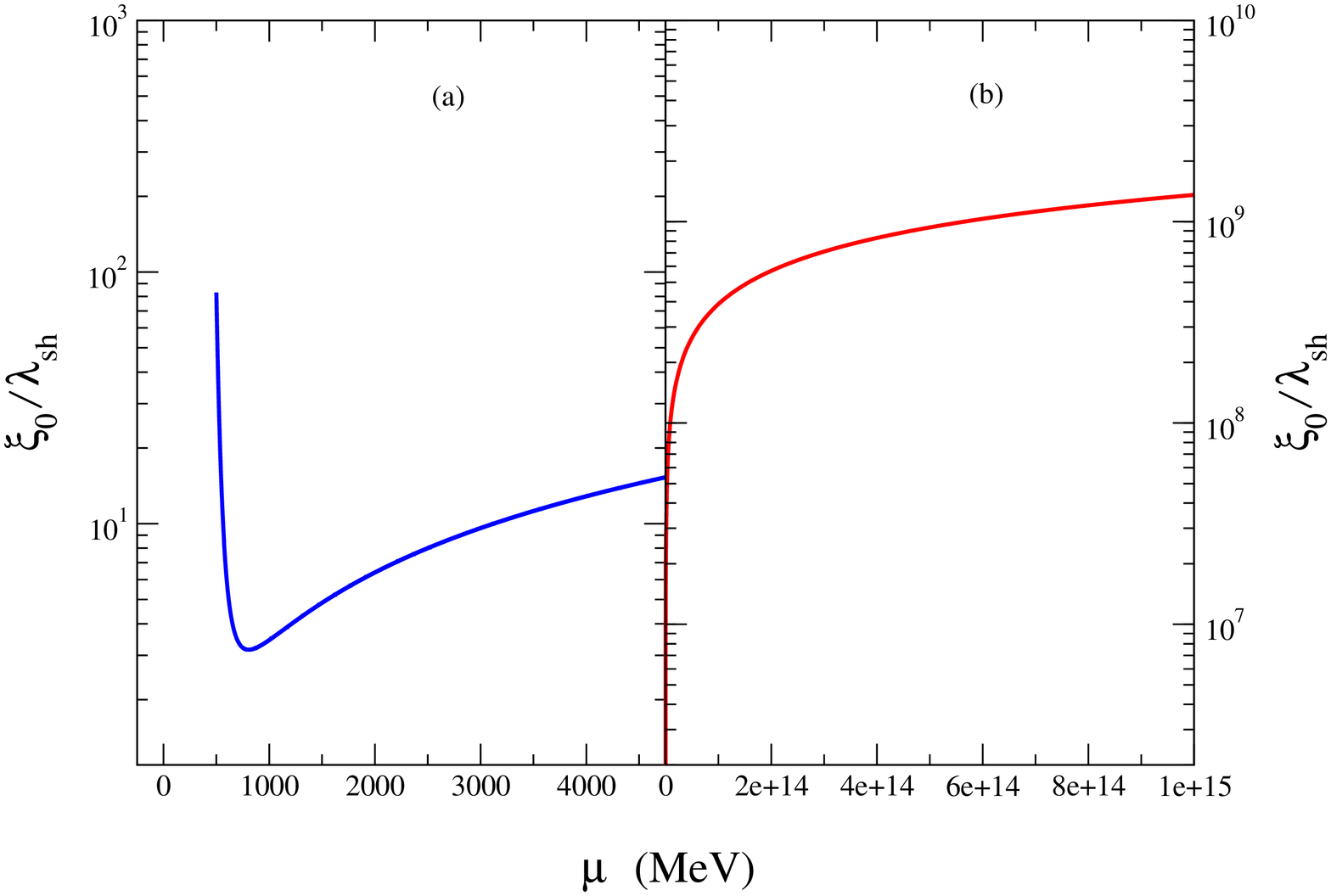}
\caption{(a) $\xi_0/\lambda_{\rm sh}$ at small chemical
potentials. (b) The same ratio at very large chemical potentials.}
\label{fig4}
\end{figure}

Now we will address the question of whether or not the London limit
is realized near $T_c^*$ for color superconductors in the range of
chemical potentials explored here. From Fig.\ \ref{fig3} we see that
the ratio $\xi_0/\lambda_{c^*} \gg 1$, meaning that only the
Pippard limit of magnetic interactions is present in color
superconductivity. Even for the minimum value of the ratio
$\xi_0/\lambda_{c^*}$, which is around $\mu=700$ MeV, the Pippard
expansion of $m^2(k,T)$ in Eq.\ (\ref{pippard1}) works better than
the London expansion, displayed in Eq.\ (\ref{london1}). This is
also the case for the metastable CSC state up to $T_{\rm sh}$, as is
shown in Fig.\ \ref{fig4}.


%

It is instructive to express the right-hand side of Eq.\
(\ref{eqvrema}) in terms of the GL parameter (\ref{glp}) and compare
it with the corresponding equation for a metallic superconductor,
whose generalized GL free energy was given in Ref.\
\cite{dirkfluct}. We have
\begin{equation}
{\cal
F}\Big(\frac{\xi_0^2}{\lambda^2_{c^*}}\Big)=\frac{3\pi^2\kappa^2}{16\alpha_s}\;,
\label{QCDcomp}
\end{equation}
for color superconductors and
\begin{equation}
{\cal
F}\Big(\frac{\xi_0^2}{\lambda^2_{c^*}}\Big)=\frac{\pi^2\kappa^2}{16
\alpha_ev_F}\;,
\label{QEDcomp}
\end{equation}
for electronic superconductivity. A large value on the right-hand
side of Eq.\ (\ref{QCDcomp}) or Eq.\ (\ref{QEDcomp}) points to the
London limit at the first-order phase transition. Since
$\alpha_e\ll\alpha_s$ and $v_F\sim\alpha_e$, the right-hand side of
Eq.\ (\ref{QEDcomp}) is much larger than that of Eq.\
(\ref{QCDcomp}) under the same GL parameter. In other words, the
London limit is more likely to be realized in metallic
superconductors.

\section { Concluding remarks }\label{V}

In this paper we have systematically calculated the effects of
gauge-field fluctuations on the free energy of a homogeneous CFL
color superconductor in the Hartree-Fock approximation. We evaluated
both analytically and numerically the temperature of the
fluctuation-induced first-order phase transition, the latent heat,
as well as the maximum temperature of a superheated superphase. It
was also shown that the London limit for color-magnetic
interactions in CFL color superconductors is absent.
As the main reason we identified the weakness of
electromagnetic interactions in comparison to strong interactions,
$\alpha_e \ll \alpha_s$. Thus, one can say that once the
gauge-field fluctuations are taken into account, the local-coupling
approximation between the order parameter and the gauge fields is
not valid in the CFL phase.

By using an inhomogeneous GL theory, Iida and Baym
\cite{iidavortices} investigated the formation of vortices and
supercurrents induced by external magnetic fields and rotation in
pairing states near the critical temperature. Since they used a
mean-field approximation, all gauge fields were regarded as averaged
quantities and fluctuations around their mean values were not
considered. In order to see how the inclusion of fluctuations would
change their results one has to derive an effective action that
depends only on the order parameter and the gauge fields. This
action would display non-local interactions between the gauge fields
and the diquark condensate. Such an effective action could be
obtained using the formalism developed in Ref.\ \cite{philipp}.

It was shown in Ref.\ \cite{ma}, by a one-loop renormalization-group
calculation using the $\epsilon$ expansion, that no stable infrared
fixed point can exist for a theory involving local interactions
between abelian gauge fields and order parameters, unless the number
of order parameter components, $N$, is artificially extended to
$N>N_{c}=365$, which is far beyond the case of relevance for
electronic superconductivity. This is then interpreted as signaling
the presence of a first-order transition. Therefore, for electronic
superconductors, gauge-field fluctuations are always expected to
change the order of the phase transition to first order,
irrespective of further details about the transition. For color
superconductors the effective action containing only
the order parameter and the gauge fields as well as
the specific form of their interactions is not known, and
the general result derived
in Ref.\ \cite{ma} may not be applicable. However, the results we
obtained for the CFL phase seem to suggest that fluctuation-induced
first-order phase transitions are indeed present in color
superconductivity. Furthermore, due to the absence of the London limit,
we expect that, once gauge-field fluctuations and
first-order phase transitions are taken into account, local
diquark-gluon interactions are never realized in color
superconductors, regardless which phase is considered. This would
constitute a striking new physical effect that would only come about
in color superconductivity. In fact, the crossover from nonlocal to
local interactions near the critical temperature in superconducting
metals of strong type I has been recently observed \cite{bonalde}.
What we found in this paper rules out the possibility of observing
such a crossover in color superconductors.

Recently, a GL free energy that takes into account the effects of
nonzero quark masses and charge neutrality has been derived within
the mean-field approximation \cite{Iidamass}. A study on the
validity of local diquark-gluon interactions in this case and the
effects of gauge-field fluctuations on the phase diagram obtained in
Ref.\ \cite{Iidamass} is in progress and will be reported elsewhere
\cite{future}.

\section*{Acknowledgments} 

The authors thank H.-J.\ Drescher, T.\ Hatsuda, J.\ Hostler, K.\ Iida, H.\
Malekzadeh, P.\ Reuter, and I.\ Shovkovy for insightful discussions.
J.L.N.\ acknowledges support by the Frankfurt International
Graduate School for Science (FIGSS) and the Volkswagen Stiftung 
and thanks the Physics Department at Rockefeller
University for its kind hospitality during a visit where part of
this work was done. The work of I.G.\ and H.C.R.\ was
supported in part by the US Department of Energy under grants
DE-FG02-91ER40651-TASKB. The work of D.H.\ was partly
supported by the NSFC under grant No.\ 10135030 and the Educational
Committee of China under grant No.\ 704035. H.C.R.\ and D.H.\
were also supported by the NSFC under grant No.\ 10575043.

\appendix

\section{}

In this appendix, we sketch some important steps for the derivation
of the generalized GL free energy in the presence of gauge-field
fluctuations, which is shown in Eq.\ (\ref{ggl}), in terms of
Feynman diagrams. We have
\begin{subequations}
\begin{eqnarray}
\bigl [ S^{-1} (P)\bigr ]^{c_1 c_2}_{f_1f_2} &=& 
(i\gamma_\mu P_\mu-\mu\gamma_4 \rho_3)\delta^{c_1 c_2}\delta_{f_1f_2}\;,
\label{A4}\\
\bigl [ \delta {\cal S}^{-1} (P)\bigr ] ^{c_1 c_2}_{f_1f_2} &=&
i\phi(P) \gamma_5 \rho_2 (\delta^{c_1}_{ f_1}\delta^{c_2}_{ f_2}
-\delta^{c_1}_{ f_2}\delta^{c_2}_{ f_1})\;,
\label{A5}\\
{\cal D}_n^{ll'}(K)_{ij}&\approx & \frac {{\delta}^{ll'}}{k^2+
\frac{\pi}{4} m_D^2 \frac{|  \omega|}{k}} \left(\delta_{ij}
-\frac{k_i k_j}{k^2}\right)\;,
\label{A6}\\
{\cal D}_n^{ll'}(K)_{j4}&=& 0 \;,
\label{A7}\\
{\cal D}_n^{ll'}(K)_{44}&\approx & \frac {\delta^{ll'}}{k^2+
m_D^2}, \label{A8}
\end{eqnarray}
\end{subequations}
where $P=(\nu, \vec p)$, $K=(\omega, \vec k)$, $
m_D^2=3\,g^2 \mu^2/(2 \pi^2)$, $\rho$ represents Pauli matrices
with respect to Nambu-Gorkov indices, $c_i$ and $l,l'$ stand for the
fundamental and adjoint color indices, respectively, $f_i$ are
fundamental flavor indices, and $\nu$, $\mu$ correspond to discrete
Matsubara frequencies. The symbol ``$\approx$" in the gluon
propagator means that we used the approximation for the total HDL
gluon propagator that is relevant for the CSC energy scale.

Diagrammatically, ${\cal  D}_n$ is denoted by a wavy line, ${\cal
S}_n$ is represented by a thick line, and the CSC correction to the
inverse quark propagator (A1b) is associated with a two-point vertex
bearing a cross. The corresponding diagrammatic expansions for
$\delta {\cal S}$ and $\delta \Pi$ are
\begin{center}
\begin{minipage}{12cm}
\vspace*{2.5cm}
\mbox{\includegraphics[width=15.0cm]{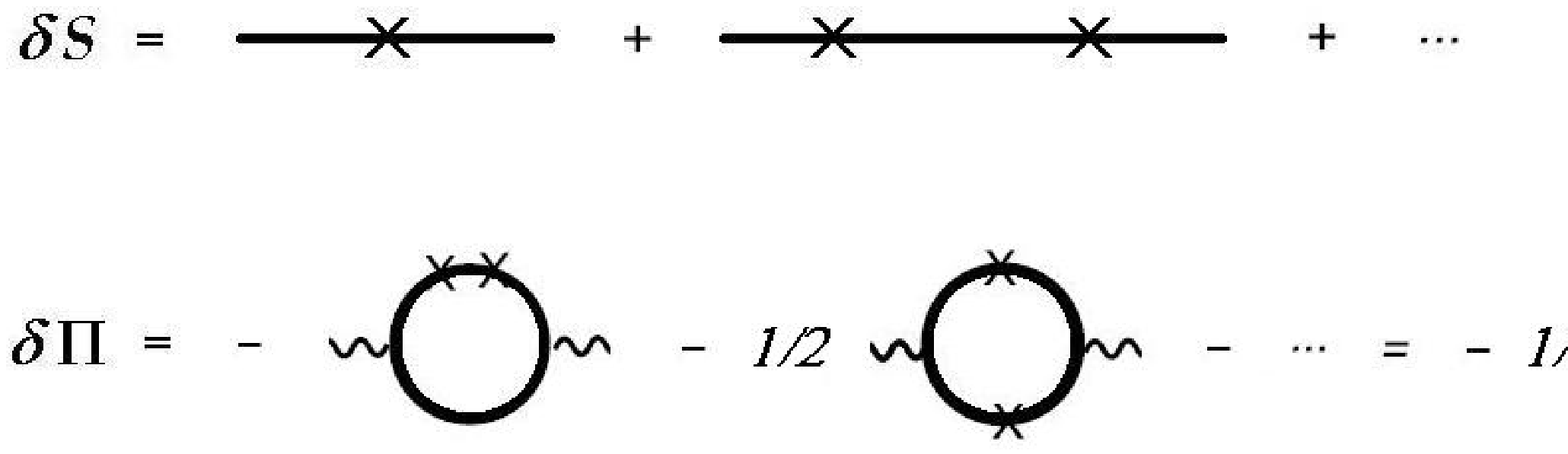}}
\end{minipage}
\end{center}
\vspace*{-2.0cm}

In weak coupling we expand ${\cal S}_n$ as
\begin{center}
\begin{minipage}{12cm}
\vspace*{3.0cm}
\mbox{\includegraphics[width=15.0cm]{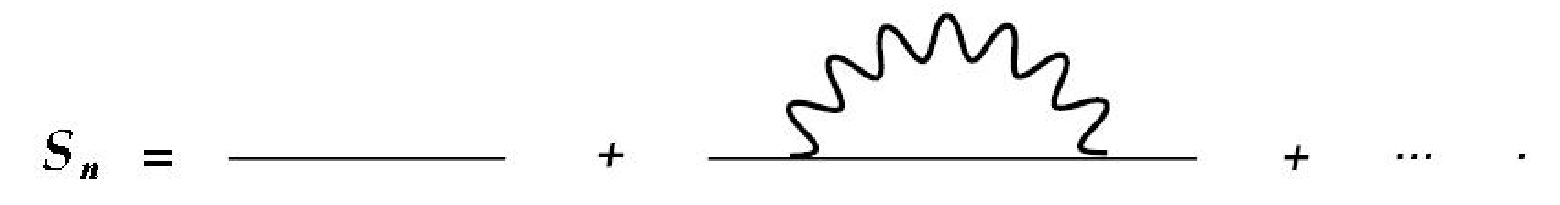}}
\end{minipage}
\end{center}
\newpage
Expanding $\,\Gamma_{\rm cond}$ up to the fourth power of
$\phi(P)$ we find

\begin{center}
\begin{minipage}{12.6cm}
\vspace*{2.0cm}
\mbox{\includegraphics[width=15cm]{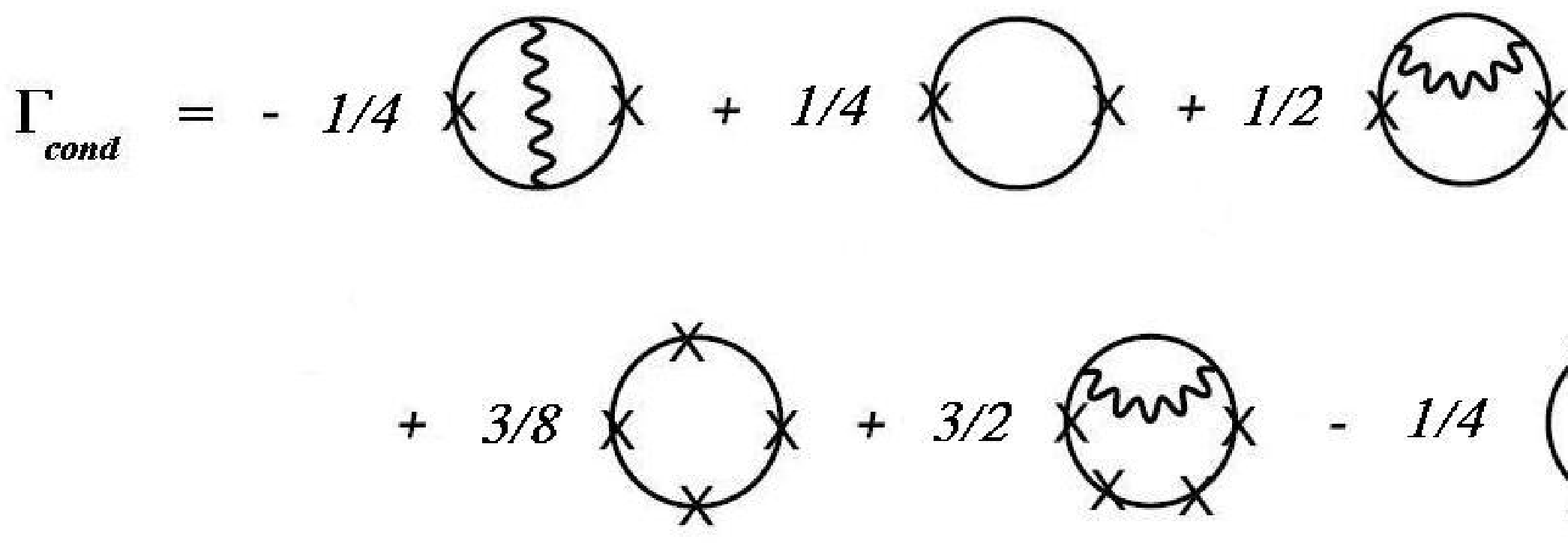}}
\end{minipage}
\end{center}
\vspace{-2.5cm}
\noindent
where the weak-coupling approximation has been
employed in order to retain the diagrams with at most one HDL gluon
line. The diagram bearing two crosses yields the expression ${\sum
_{PP'} }\Phi(P) K(P|P') \Phi(P')$, where the kernel $K(P|P')$ is
isomorphic to the kernel in the Dyson-Schwinger equation for the
diquark scattering amplitude in the normal phase. Moreover, taking
$\Phi(P)$ to be proportional to the pairing mode, i.e.,
\begin{equation}
\phi(P)=\Delta\sin\Big[\,\frac{g}{3\sqrt{2}\pi}\ln\Big(\frac{1}{\hat\nu}\Big)
\Big]\hspace{0.3cm},
\end{equation}
where $\Delta$ is the energy gap, $\hat \nu
=(3/2)^{5/2}\, g^5 \nu/(256 \pi^4\mu)$, we
have that

\begin{center}
\begin{minipage}{8.0cm}
\vspace*{3cm}
\includegraphics[width=8cm]{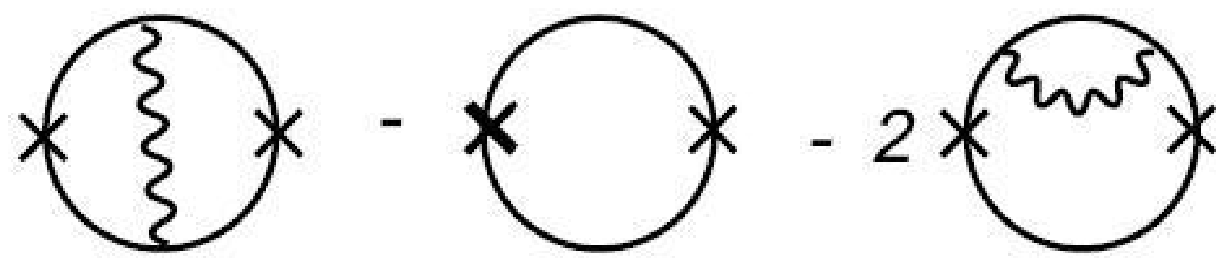}
\end{minipage}
\end{center}
\vspace{-3.0cm}
\noindent is proportional to $T-T_c$, with $T_c$ determined up
to subleading order in $g$ [see Eq.\ (\ref{pairing})]. For the
diagrams with four crosses the same mechanism yields

\begin{center}
\begin{minipage}{9.0cm}
\includegraphics[width=6cm]{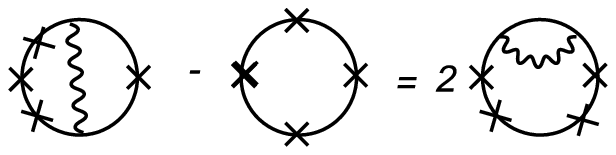}
\end{minipage}
\end{center}
\noindent at $T=T_c$, which reduces the number of quartic
terms in $\Gamma_{\rm cond}$. Moreover, it will be shown at the end of
this appendix that the following two diagrams

\begin{center}
\begin{minipage}{17.6cm}
\vspace*{-0.5cm}
\begin{equation}
\includegraphics[width=6cm]{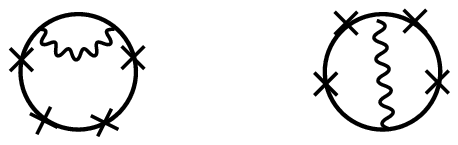}
\label{A9}
\end{equation}
\end{minipage}
\end{center}
\noindent
are of higher order in weak coupling and can be
dropped. For $\, \Gamma_{\rm cond}$ we end up with


\begin{center}
\begin{minipage}{10cm}
\vspace*{2.3cm}
\includegraphics[width=14cm]{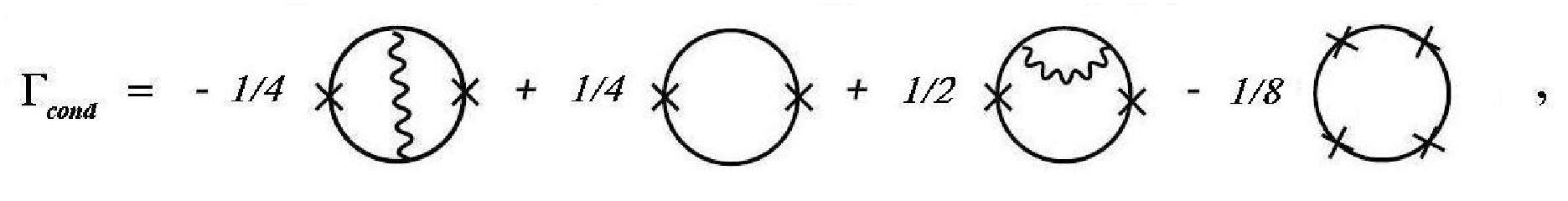}
\end{minipage}
\end{center}
\vspace{-2cm}
\noindent
which produces the terms in Eq.\ (\ref{free}). Now
we treat the fluctuation terms $ \Gamma_{\rm fluc}+\Gamma'_{\rm fluc}$ ,
whose diagrammatic representation is
\begin{center}
\begin{minipage}{17.6cm}
\vspace*{3cm} \hspace*{2cm}
\includegraphics[width=16cm]{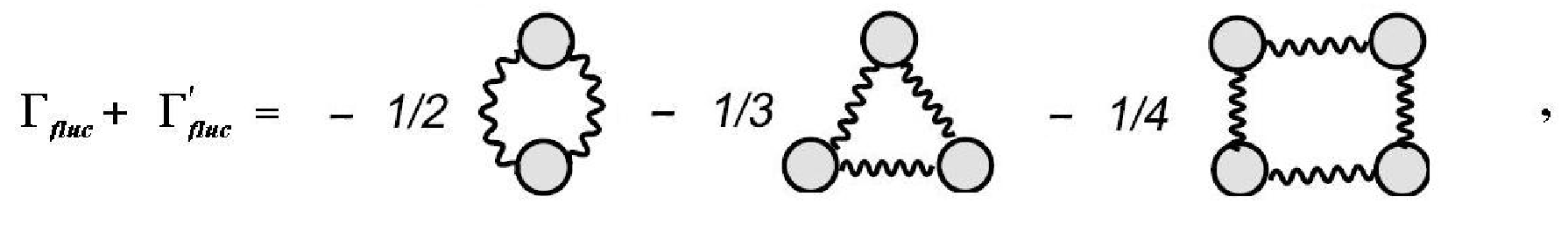}
\vspace*{-5cm}
\begin{equation}
\,
\label{A11}
\end{equation}
\end{minipage}
\end{center}
\noindent where $\Gamma_{\rm fluc}$ includes only the contribution
from the static gluons and $\Gamma'_{\rm fluc}$ contains remaining
contributions. Due to the Meissner effect, the shaded bubble does
not vanish when the spatial momentum of the gluon line goes to zero
at zero Matsubara energy. A resummation of all ring diagrams in Eq.\
(\ref{A11}) is necessary for $\Gamma_{\rm fluc}$ and the result is the
right-hand side of Eq.\ (\ref{fluct}). Regarding $\Gamma'_{\rm fluc}$,
where the Matsubara energy of the gluon line is nonzero, dynamical
screening prevents an infrared divergence for the integral over
gluon momentum. In weak coupling $\Gamma'_{\rm fluc}$ is dominated  by
the first diagram in Eq.\ (\ref{A11}), which is again of higher
order. Therefore, the contribution of $\Gamma'_{\rm fluc}$ can be safely
neglected.

Now we present the argument supporting our assertion that the two
diagrams in Eq.\ (\ref{A9}) and the first diagram in
$\Gamma'_{\rm fluc}$ can be neglected in weak coupling. Let us denote
the contribution of the first diagram in Eq.\ (\ref{A9}) by $c_1
\Delta^4$. It consists of five free quark propagators with
four-momentum 
$(\nu, \vec{p})$ and a self-energy insertion $\Sigma(P)
\sim g^2 \nu \ln\left(\mu/|\nu|\right) \sim g \nu$. The main
contribution to the $\vec p$-integration comes from a shell of
thickness $\sim |\nu|$ around the Fermi surface and then we have

\begin{equation}
c_1\sim  \mu^2 T {\sum_\nu} \frac {1} {| \nu|^4} \, \Sigma (\nu)
\sim g\,\frac {\mu^2}{T_c^2}, \label{A19}
\end{equation}

\hspace{-0.35cm}which is of ${\cal O} (g)$ in comparison to the
quartic term in Eq.\ (\ref{free}). The contribution of the second
diagram in Eq.\ (\ref{A9}), denoted by $c_2 \Delta ^4$, can be
estimated similarly. As is the case with the gap equation, the
dominating contribution comes from the magnetic gluons with nonzero
Matsubara energy. The integration for the quark propagators over the
magnitude of their momenta $\vec p$, $\vec{p}$ $'$, on each side of
the gluon line can be approximately decoupled from the integration
for the gluon propagator over the angle between $\vec p$ and
$\vec{p}$ $'$, where the latter produces the forward logarithm. We
then find

\begin{equation}
c_2\sim  g^2 T^2 \mu^4  {\sum_{\nu\not = \nu'}} \frac {1} {\nu^2
\nu'^2} \, \frac{1}{\mu^2} \ln \left(\frac{\mu}{\nu -\nu'}\right)
\sim g\,\frac {\mu^2}{T_c^2}, \label {A20}
\end{equation}
which is again of higher order. Now we consider the first diagram in
$\Gamma'_{\rm fluc}$ and denote its contribution as $c_3 \Delta ^4$.
Since the typical momentum for the gluon line is $k\sim
m_D^{2/3} |\omega|^{1/3}\gg \omega $ and $\omega \sim
T_c$, each bubble can be approximated by the static magnetic
self-energy of gluons at the Pippard limit, i.e.,

\begin{equation}
c_3\sim  g^4 T   {\sum_{\omega\not = 0}} \int\frac{d^3\vec
k}{(2\pi)^3}\, \frac {1}{(k^2+ \frac{\pi}{4} m_D^2 \frac{|
\omega|}{k})^2}  \left(\frac{\mu^2}{T_c^2}\, \frac{T_c}{k}\right)^2
\sim \frac{g^4 \mu^4}{T_c^2 m_D^2} {\sum_{\omega\not = 0} } \frac
{1}{|\omega|}\, . \label {A21}
\end{equation}
The sum over $\omega$ has a cutoff when $\omega\sim m_D$ and then we
end up with ${\sum_{\omega\not = 0} } |\omega|^{-1}\sim
\ln\left(\mu/T_c\right) \sim 1/g$. Consequently, we
have $c_3 \sim g\,\mu^2/T_c^2$, which is also negligible.


\end{document}